\documentclass[aps,pra,twocolumn,superscriptaddress]{revtex4}
\usepackage{graphicx}
\usepackage{dcolumn}

\begin{document}

\title{Anomalous dynamical evolution and nonadiabatic level crossing in exactly
solvable time-dependent quantum systems}

\author{Hong Cao}
\affiliation{Center of Theoretical Physics, College of Physics, Sichuan University, Chengdu 610065, China}

\affiliation{School of Material Science and Engineering, Chongqing Jiaotong
University, Chongqing 400074, China}

\author{Shao-Wu Yao}
\affiliation{Center of Theoretical Physics, College of Physics, Sichuan University, Chengdu 610065, China}

\author{Li-Xiang Cen}
\email{lixiangcen@scu.edu.cn}
\affiliation{Center of Theoretical Physics, College of Physics, Sichuan University, Chengdu 610065, China}

\begin{abstract}
The anomalous dynamical evolution and the crossing of nonadiabatic
energy levels are investigated for exactly solvable time-dependent
quantum systems through a reverse-engineering scheme. By exploiting a typical driven model,
we elucidate the peculiarities of its dynamics with anomalous
behavior: the evolution of the adiabatic states and of the nonadiabatic ones exhibits opposite
behavior with their representative vectors evolving from a
parallel state to an antiparallel state;
the nonadiabatic level crossing is
identified as a necessary consequence since the crossing point corresponds
exactly to the perpendicular point of the two vectors in the parametric space.
In the light of these results, we show that
various driven models with anomalous dynamical evolution can be designed
and they offer alternative protocols for the quantum state control.
\end{abstract}

\maketitle

\section{Introduction}

Nonadiabatic dynamics generated by quantum systems with explicitly
time-dependent Hamiltonians plays an important role in various branches of
quantum physics. Especially, when the diabatic energies of two interacting
quantum states are forced to cross, e.g., by a linearly driving external
field, the nonadiabatic transition between adiabatic levels is known as the
Landau-Zener (LZ) tunneling \cite{LZ1,LZ2}. The latter offers a simple way
to understand the wave phenomenon of the quantum system and the resulting
quantitative formula of the transition probability has widespread
applications ranging from quantum optics and atomic physics \cite
{opt1,opt2,opt3,opt4,opt5} to chemistry and biophysics \cite
{chem1,chem2,chem3,chem4}. Recently, application of the LZ driving and its
generalized protocols to the quantum state manipulation has attracted much
attention in the context of quantum information processing \cite
{gener,gener0,gener1,gener3,gener5,gener6,gener7}. Experimental
demonstration of the corresponding coherent dynamical transition has been
reported in a variety of physical systems, e.g., the Rydberg atoms \cite{exper0},
the superconducting quantum interference device \cite{exper1,exper2},
and the nitrogen-vacancy center in diamond \cite{exper3}.

The evolution of the energy level and the wavefunction of a driven quantum
system can be distinctly different under the adiabatic or the nonadiabatic
driving. In the case of the adiabatic evolution, the LZ-type driven model
will exhibit the avoided level crossing which is
well understood. Consider as an example the dynamics generated by a
two-level system with the following Hamiltonian
\begin{equation}
H(t)=\frac 12\left(
\begin{array}{ll}
\Omega _z(t) & \Omega _x(t) \\
\Omega _x(t) & -\Omega _z(t)
\end{array}
\right),  \label{hamil0}
\end{equation}
in which the bare energies $\pm\frac 12\Omega_z (t)$ change from $\Omega_z<0
$ to $\Omega_z >0$ at some time instant, e.g., at $t=0$. The occurrence of a
perturbative $\Omega_x(t)$ around $t=0$ will lift the degeneracy of the
energies as $E_{\pm}^{ad}(t)= \pm\frac 12 \sqrt{\Omega_x^2(t)+\Omega_z^2(t)}$,
which results in the emergence of the avoided level crossing (the
particular case with both $\Omega_x(t)$ and $\Omega_z(t)$ vanishing at $t=0$
was investigated in Ref. \cite{realc}). The adiabatic following of the
corresponding instantaneous eigenstate of the Hamiltonian then leads to the
population transfer from one bare state to the other. On the other hand, as
the nonadiabatic driving is concerned, the dynamics generated by this kind
of models depends heavily on the pulse shape of the driving field.
For example, in the standard LZ protocol with linearly driving field,
the nonadiabaticity-induced transition will destroy the desired population
transfer. Notwithstanding, recent studies on the variants of the LZ model
display that the complete population transfer could be achieved through the
avoided level crossing under a tangent-shape driving \cite{gener6} even
when the corresponding evolution is in a nonadiabatic manner.

In the general case of the nonadiabatic evolution, the Hamiltonian itself is
no longer an invariant of the system. The energy levels are then defined by
the expectation values of the Hamiltonian over the nonadiabatic bases, e.g.,
the eigenvectors of the Lewis-Riesenfeld (LR) invariant \cite{lewis,lewis2}.
The phenomenon of the nonadiabatic level crossing (NLC), if it does occur,
would have nothing to do with the degeneracy or the symmetry of the system. So the
question arises: what is the dynamical implication with respect to the
real NLC? This issue had hardly been investigated, probably because of
the fact that the case of driven quantum systems that could exhibit the
NLC phenomenon is very rare. In this paper we shall explore this issue
in virtue of exactly solvable time-dependent quantum systems that are constructed
through a reverse-engineering scheme. We will first describe the anomalous dynamical
behavior in a nonadiabatically driven system, that is, the nonadiabatic state of the system
exhibits opposite behavior from its adiabatic state. We then demonstrate that the occurrence
of the NLC constitutes a necessity condition for such particular dynamical behavior.
The general description of the driving protocol and illustration of
various examples with distinct features will be presented within the
reverse-engineering scheme.

\section{A typical driven model with anomalous dynamical evolution}

\begin{figure}
\resizebox{0.45\textwidth}{!}{
\includegraphics{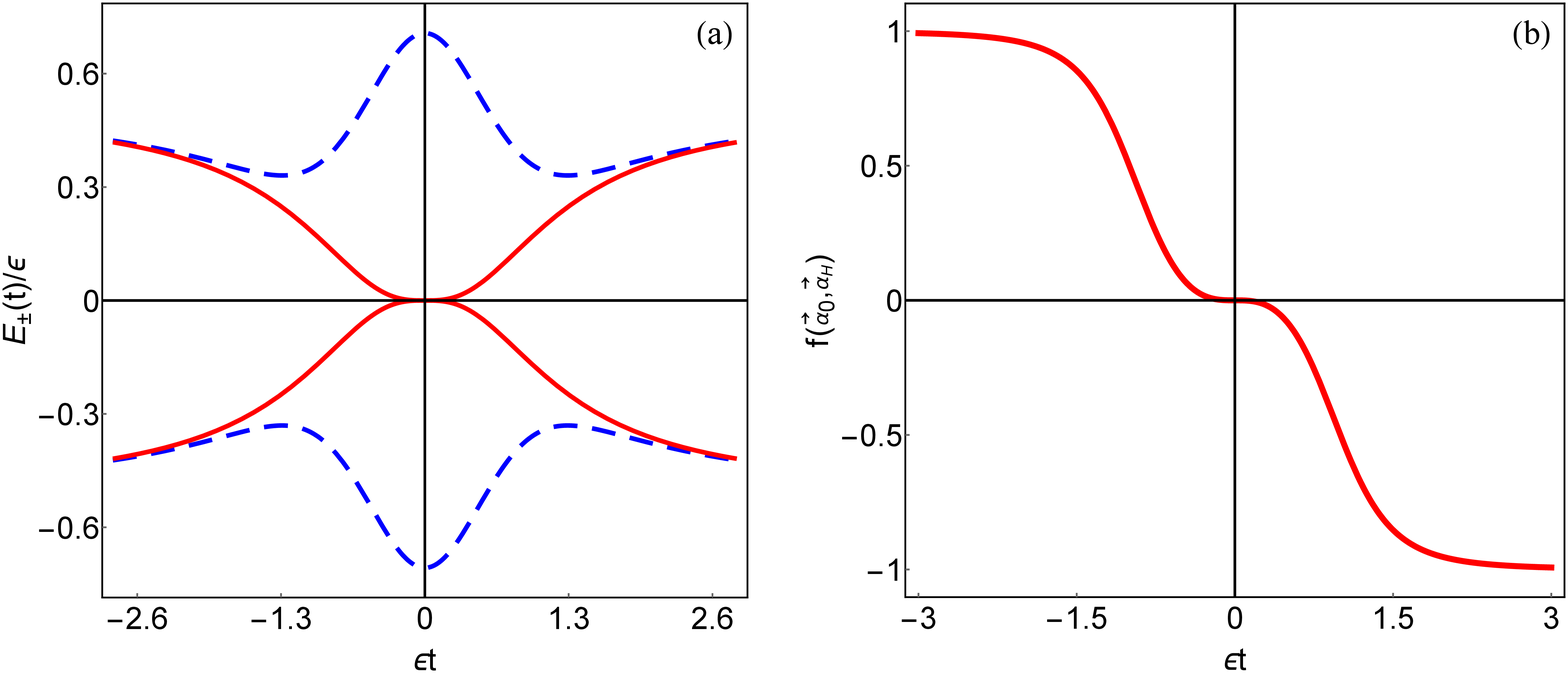}}
\caption{The NLC phenomenon of the model (\ref{hamil1}) with $j=\frac 12$.
(a) The two nonadiabatic energy levels $E_{\pm}(t)$ over $\epsilon$ which intersect at $t_c=0$.
The dashed lines stand for the instantaneous adiabatic energies $E_\pm^{ad}(t)$
over $\epsilon$ which exhibit the avoided level crossing twice. (b) The inner product
$f(\vec{\alpha}_0,\vec{\alpha}_h)$ changes from $+1$ to $-1$ and vanishes at $t_c=0$.}
\label{figNLC}
\end{figure}

Let us firstly consider a driven quantum system described by the following
Hamiltonian
\begin{eqnarray}
H(t) &=&\Omega_x(t)J_x+\Omega_z(t)J_z  \nonumber \\
&=&\frac \epsilon {\sqrt{\epsilon ^2t^2+1}}J_x+\frac{\epsilon (\epsilon
^2t^2-1)}{\epsilon ^2t^2+1}J_z,  \label{hamil1}
\end{eqnarray}
where $J_i$ $(i=x,y,z)$ denote the angular momentum operators satisfying
$[J_i,J_j]=i\varepsilon _{ijk}J_k$ and the driving field has two components
along the $x$ and $z$ axes, respectively. The orientation of the driving
field, specified by $\vec{\alpha}_h(t)=\frac {\vec{\Omega}(t)}{|\vec{\Omega}
(t)|}$, tends to $+z$ axis at $t\rightarrow \pm \infty $. It indicates that
the instantaneous adiabatic eigenstates of $H(t)$, expressed as $|\psi
_m^{ad}(t)\rangle =e^{-i\theta _h(t)J_y}|m\rangle $ with $|m\rangle$ denoting
the eigenstate of $J_z$  and
$\theta_h(t)=\arccos \frac{\Omega _z(t)}{\sqrt{\Omega _x^2(t)+\Omega _z^2(t)}}$,
will return to its initial state
$|\psi _m^{ad}(-\infty )\rangle =|m\rangle $ at $t\rightarrow +\infty $. This
can also be understood from the fact that the adiabatic energy levels $
E_m^{ad}(t)$ undergo the avoided crossing twice during the evolution [see
Fig. \ref{figNLC} (a)].

To resolve the nonadiabatic evolution of the system governed by the
Schr\"{o}dinger equation (setting $\hbar =1$)
\begin{equation}
i\frac \partial {\partial t}|\psi (t)\rangle =H(t)|\psi (t)\rangle ,
\label{schrod}
\end{equation}
it is direct to verify that the system possesses a dynamical invariant
\begin{equation}
I(t)=-\frac{\epsilon t}{\epsilon ^2t^2+1}J_x-\frac 1{\epsilon ^2t^2+1}J_y-
\frac{\epsilon t}{\sqrt{\epsilon ^2t^2+1}}J_z  \label{invar}
\end{equation}
which satisfies $i\partial _tI(t)=[H(t),I(t)]$. According to the LR theory \cite{lewis,lewis2},
the solution to the Schr\"{o}dinger equation can be achieved by the
instantaneous eigenstate of $I(t)$ equipped with a phase factor.
Let us express $I(t)$ as $I(t)=\vec{\alpha}_0(t)\cdot \vec{J}$ in which
$\vec{\alpha}_0(t)=(\sin \theta _0\cos \varphi _0,\sin \theta _0\sin \varphi
_0,\cos \theta _0)$ and $\theta _0(t)$ and $\varphi _0(t)$ are given by
\begin{equation}
\theta _0(t)=\frac \pi 2+\arctan (\epsilon t),~~\varphi _0(t)=\frac{3\pi }
2-\arctan (\epsilon t).  \label{angle}
\end{equation}
The eigenstate of $I(t)$ can then be obtained explicitly as $|\phi _m(t)\rangle
=e^{i(\pi -\varphi _0)J_z}e^{i\theta _0J_y}|m\rangle $. So the dynamical basis
of the system $|\psi _m(t)\rangle$ is formulated as $|\psi _m(t)\rangle
=e^{i\Phi _m(t,t_0)}|\phi _m(t)\rangle $, where
$\Phi _m(t,t_0)$ is the so-called LR total phase given by
\begin{eqnarray}
\Phi _m(t,t_0) &=&\int_{t_0}^t\langle \phi _m(t^{\prime })|i\partial
_{t^{\prime }}-H(t^{\prime })|\phi _m(t^{\prime })\rangle dt^{\prime }
\nonumber \\
&=&-m\int_{t_0}^t\frac{\epsilon ^2\tau }{\sqrt{\epsilon ^2\tau ^2+1}}d\tau .
\label{totph}
\end{eqnarray}

With the above analytical solution, we are able to describe the anomaly of
the dynamical evolution of the model. It is seen that as $t$ goes from
$-\infty $ to $+\infty $, the invariant $I(t)$ will go from $+J_z$ to $-J_z$.
Indeed, the two representative vectors $\vec{\alpha}_h(t)$ and $\vec{\alpha}_0(t)$,
accounting respectively for the
orientation of $H(t)$ and $I(t)$ in the parametric space, will evolve from
the parallel state with $\theta _h=\theta _0=0$ to the antiparallel state with
$\theta _h=0$ and $\theta _0=\pi $.
Therefore, the nonadiabatic basis state $|\psi _m(t)\rangle $ exhibits the
opposite behavior from the adiabatic state $|\psi _m^{ad}(t)\rangle $ as it
leads to a complete population inversion $|m\rangle \leftrightarrow
|-m\rangle $ during the evolution.
Furthermore, the peculiarity of the dynamical behavior can also be recognized from
the different crossing phenomena of the corresponding adiabatic
and nonadiabatic energy levels. As the adiabatic levels exhibit only avoided crossings, the
nonadiabatic levels of the model, formulated as
\begin{eqnarray}
E_m(t) &\equiv &\langle \psi _m(t)|H(t)|\psi _m(t)\rangle   \nonumber \\
&=&-\frac{m\epsilon ^4t^3}{(1+\epsilon ^2t^2)^{3/2}},  \label{nonlevel}
\end{eqnarray}
exhibit a real crossing at $t_c=0$. For the two-level case with
the azimuthal quantum number $j=\frac 12$, these level crossing phenomena are
illustrated in Fig. \ref{figNLC}(a).

\section{General description of the protocol}

\subsection{Condition associated with the nonadiabatic level crossing}

To reveal the connection between the above
described anomalous dynamical evolution and the NLC phenomenon,
it is worthy to note that at the crossing point $t_c=0$ the orientation of
$\vec{\alpha}_h$ is vertical to $\vec{\alpha}_0$
as $\vec{\alpha}_h$ is fixed in the $x$-$z$ plane and
$\vec{\alpha}_0(t_c)=(0,-1,0)$ is exactly along the $y$ axis.
In fact, this is not accidental in view that the nonadiabatic level
$E_m(t)$ can be expressed as
\begin{eqnarray}
E_m(t)&=&\langle \phi _m(t)|H(t)|\phi _m(t)\rangle \nonumber \\
&=&\langle m|e^{-i\theta_0J_y}e^{-i(\pi-\varphi_0)J_z}H(t)e^{i(\pi-\varphi_0)J_z}e^{i\theta_0J_y}|m\rangle \nonumber \\
&=&m\vec{\Omega}(t)\cdot \vec{\alpha}_0(t).
\label{levelg}
\end{eqnarray}
That is to say, there exists a general correspondence between the orthogonal
relation $\vec{\alpha}_0\cdot\vec{\alpha}_h=0$ and the NLC
with all $E_m=0$. The evolution of the inner product
$f(\vec{\alpha}_0,\vec{\alpha}_h)\equiv \vec{\alpha}_0(t)\cdot\vec{\alpha}_h(t)$ of the model
(\ref{hamil1}) is illustrated in Fig. \ref{figNLC}(b). Mathematically, the
orthogonality here can also be understood as that the two operators, $I(t)$
and $H(t)$, have vanishing Frobenius inner product \cite{froben}: $\mathrm{tr%
}[I(t_c)H(t_c)]=0$. As a consequence, the occurring
of the NLC can be identified as a necessity
condition of the anomalous dynamical evolution: as the two vectors $\vec{
\alpha}_0(t)$ and $\vec{\alpha}_h(t)$ change continuously from a parallel
state to an anti-parallel state in the parametric space, it is unavoidable
that they should go through a perpendicular position during the evolution.

\subsection{The reverse-engineering scheme}

In the light of the above result, we now address the question how to achieve
such kind of driven quantum systems with the similar anomalous dynamical
behavior. To this end, let us consider the dynamics generated by the driven
model of Eq. (\ref{hamil1}) with general field pulses $\Omega _x(t)$ and $\Omega _z(t)$.
We suppose that the system possesses a dynamical invariant
$I(t)\equiv \vec{\alpha}(t)\cdot \vec{J}$ with
its components $\alpha _i(t)$ fulfilling
\begin{eqnarray}
\dot{\alpha}_x(t) &=&-\Omega _z(t)\alpha _y(t),  \label{comp1} \\
\dot{\alpha}_y(t) &=&\Omega _z(t)\alpha _x(t)-\Omega _x(t)\alpha _z(t),
\label{comp2} \\
\dot{\alpha}_z(t) &=&\Omega _x(t)\alpha _y(t).  \label{comp3}
\end{eqnarray}
Since the eigenvalues of $I(t)$ are independent of time, we can set $|\vec{%
\alpha}(t)|=1$ and express it as $\vec{\alpha}(t)=(\sin \theta \cos \varphi
,\sin \theta \sin \varphi ,\cos \theta )$. We then regard Eqs. (\ref{comp1}%
)-(\ref{comp3}) as algebraic equations of $\Omega _{x,z}(t)$. Explicitly,
Eq. (\ref{comp3}) indicates
\begin{equation}
\Omega _x(t)=-\frac{\dot{\theta}(t)}{\sin \varphi (t)},  \label{inver1}
\end{equation}
and Eq. (\ref{comp1}) [or Eq. (\ref{comp2}) equivalently] indicates
\begin{eqnarray}
\Omega _z(t) &=&\dot{\varphi}(t)+\Omega _x(t)\cot \theta (t)\cos \varphi (t)
\nonumber \\
&=&\dot{\varphi}(t)-\dot{\theta}(t)\cot \theta (t)\cot \varphi (t).
\label{inver2}
\end{eqnarray}
That is to say, for any given $I(t)$ with specified $\theta (t)$ and $%
\varphi (t)$, one can construct the field components $\Omega _{x,z}(t)$ in
virtue of Eqs. (\ref{inver1}) and (\ref{inver2}). Note that similar
reverse-engineering schemes have ever been proposed, e.g., in Refs. \cite{berry}
and \cite{inverLR}. In comparison, there is no redundant
parameter in the present scheme and the driving field of the target Hamiltonian
here is uniquely determined. On other words, it indicates
that the driving protocol with two field components is sufficient to generate
any desired SU(2) rotation to the wavefunction.

If we set $\theta (t)=\theta _0(t)$ and $\varphi (t)=\varphi _0(t)$ specified in Eq. (\ref{angle})
and substitute them into Eqs. (\ref{inver1}) and (\ref{inver2}), one
promptly obtains the previous model of Eq. (\ref{hamil1}). To illustrate
the selection of $\theta(t)$ and $\varphi(t)$ and the resulting anomalous
dynamical behavior, one is led to notice that as $t\rightarrow \pm\infty$, there is
$\dot{\varphi}_0(t)\rightarrow 0$ and $\Omega _z$ is then dominated by
the term $\Omega _x\cot \theta _0\cos \varphi_0$ [cf. Eq. (\ref{inver2})].
So the relation between the two angles $\theta _0$ and $\theta _h$ at
$t\rightarrow \pm \infty $ is characterized by
\begin{equation}
\cot \theta _h=\cos \varphi_0 \cot \theta _0.  \label{match}
\end{equation}
The change of the relative orientation of $\vec{\alpha}_h(t)$ and $\vec{
\alpha} _0(t)$, i.e., from the parallel state to the antiparallel state, is
then well understood because $\cos \varphi _0$, according to Eq. (\ref{angle}),
alters its sign (from $+1$ to $-1$) during the evolution.
This illuminates a way to contrive similar driving protocols with the
anomalous dynamical behavior.

\subsection{Further examples of the driving protocol}

For the second example let us consider a single-axis driven model described
by
\begin{equation}
H(t)=\epsilon J_x+\Omega _z(t)J_z,  \label{hamil2}
\end{equation}
in which $\Omega_x=\epsilon $ is time independent and the driving component $\Omega _z(t)$ is
given by
\begin{equation}
\Omega _z(t)=\frac{5\epsilon e^{2\epsilon t}-\epsilon e^{6\epsilon t}}{
1+e^{4\epsilon t}}.  \label{zcomp}
\end{equation}
This model can be constructed from Eqs. (\ref{inver1}) and (\ref{inver2}) by
setting the two angles as
\begin{equation}
\theta (t)=\frac \pi 2-\arctan e^{2\epsilon t},~~\varphi (t)=2\arctan
e^{2\epsilon t}.  \label{angl}
\end{equation}
As $t$ goes from $-\infty $ to $+\infty $, these angles will go from $(\theta
,\varphi )=(\frac \pi 2,0)$ to $(\theta ,\varphi )=(0,\pi)$.
Differing from the former model, the representative vectors
$\vec{\alpha}_h(t)$ and $\vec{\alpha}(t)$ of the Hamiltonian and of
the dynamical invariant here,
are along the $+x$ axis initially at $t\rightarrow -\infty $.
They arrive eventually at an anti-parallel state,
$\vec{\alpha}_h(t)\rightarrow -z$ and $\vec{\alpha}(t)\rightarrow +z$
as $t\rightarrow +\infty $.
The anomalous dynamical behavior can be specified explicitly
by the evolution of the wavefunction. The instantaneous adiabatic states,
say, for the two-level case with azimuthal quantum number $j=\frac 12$,
will evolve as
\begin{equation}
|\psi _{\pm }^{ad}(-\infty )\rangle =\frac{\sqrt{2}}2(|+\rangle \pm
|-\rangle )\stackrel{t\rightarrow +\infty }{\longrightarrow }|\mp \rangle ,
\label{evoad}
\end{equation}
in which we have used the notation ``$\pm$" for ``$\pm\frac 12$".
In comparison, the eigenstate of $I(t)$, accounting for the
the nonadiabatic evolution of the system, will evolve as
\begin{equation}
|\phi _{\pm }(-\infty )\rangle =\frac{\sqrt{2}}2(|+\rangle \pm |-\rangle )%
\stackrel{t\rightarrow +\infty }{\longrightarrow }|\pm \rangle .  \label{evo}
\end{equation}
That is, the nonadiabaticity of the evolution induces a complete exchange between the two
adiabatic basis states as $t\rightarrow +\infty $.

\begin{figure}
\begin{center}
\resizebox{0.32\textwidth}{!}{
\includegraphics{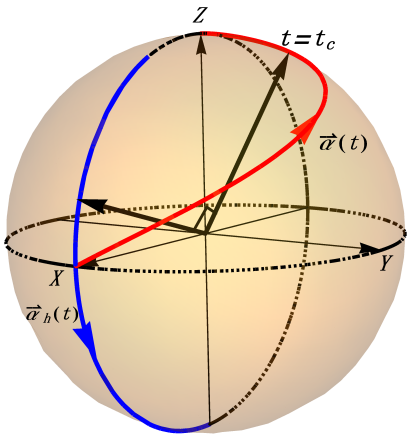}}
\end{center}
\caption{Illustration of the evolving trajectories of the two vectors $\vec{\alpha}_h(t)$
and $\vec{\alpha}(t)$ in the parametric space of the model specified by Eq. (\ref{hamil2}).
At $t=t_c$ the two vectors are shown to be vertical to each other
[cf. Eqs. (\ref{vect1}) and (\ref{vect2})].}
\label{figpath}
\end{figure}

On the other hand, the nonadiabatic energy level of the system is worked out to be
\begin{equation}
E_m(t)=m\epsilon \left[ \frac{4e^{4\epsilon t}}{(1+e^{4\epsilon t})^{3/2}}-\frac{%
e^{4\epsilon t}-1}{(1+e^{4\epsilon t})^{1/2}}\right] .  \label{dial}
\end{equation}
It is straightforward to verify that as $\epsilon t_c=\frac 14\ln (2+\sqrt{5}
)$ there is $E_m(t_c)=0$, that is, all the nonadiabatic energy levels should
intersect at the particular point $t=t_c$. Specifically, at the point of
$\epsilon t_c=\frac 14\ln (2+\sqrt{5})$, there
are
\begin{equation}
\vec{\alpha}(t_c)=(\frac{\sqrt{10}-3\sqrt{2}}4,\frac{\sqrt{2\sqrt{5}-4}}2,
\frac{\sqrt{\sqrt{5}+1}}2),  \label{vect1}
\end{equation}
and
\begin{equation}
\vec{\Omega}(t_c)/\epsilon =(1,0,\frac{\sqrt{10\sqrt{5}-22}}2).
\label{vect2}
\end{equation}
So the orthogonality of these two vectors at the crossing point $t=t_c$
can be verified straightforwardly. In Fig. \ref{figpath} we plot the schematic of the evolving
trajectories of $\vec{\alpha}(t)$ and $\vec{\alpha}_h(t)$ for the model in the parametric space.

\begin{figure}
\resizebox{0.45\textwidth}{!}{
\includegraphics{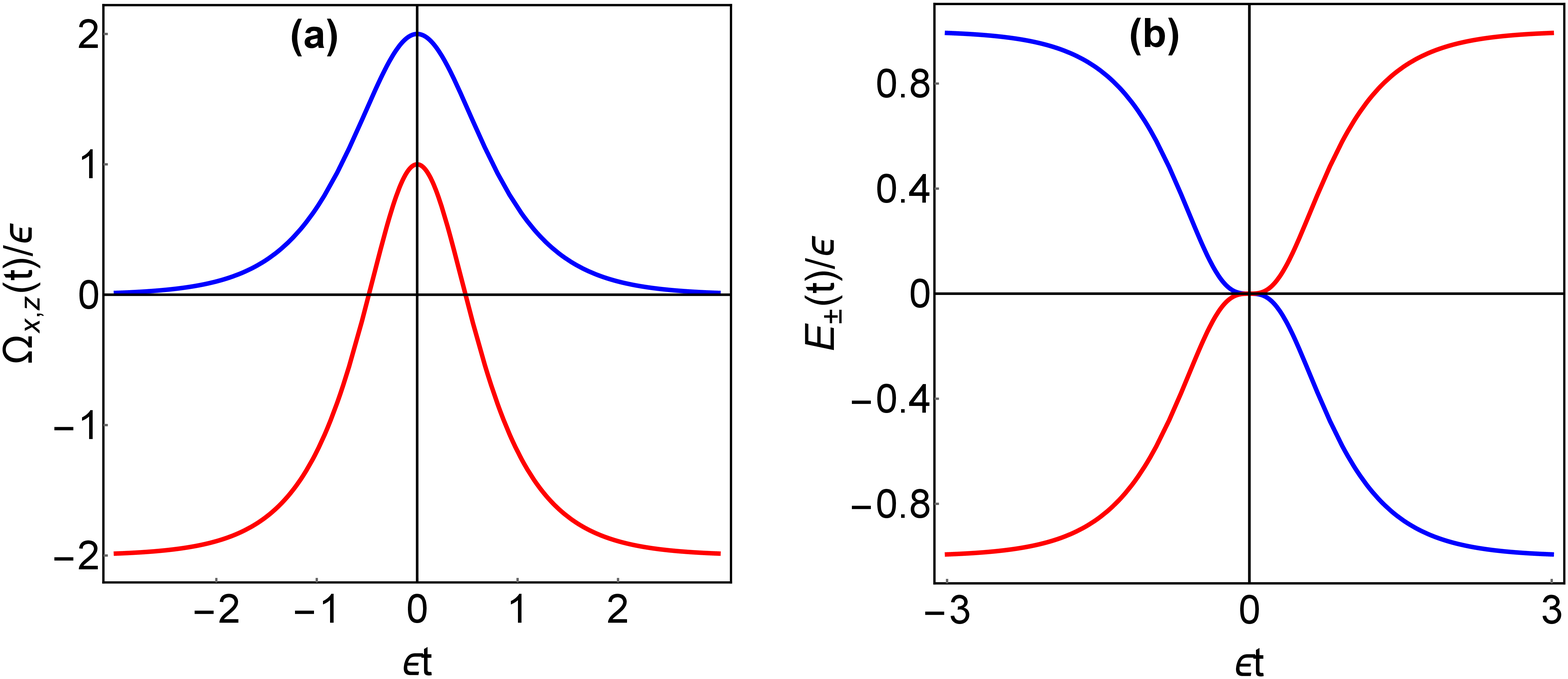}}
\caption{The field pulses $\Omega_{x,z}(t)$ and the nonadiabatic levels of
the model specified by Eq. (\ref{hamil3z}). (a) The
field component $\Omega_x(t)$ (the upper curve) vanishes as $t\rightarrow
\pm\infty$ while the component $\Omega_z(t)=-2\epsilon$ as $t\rightarrow
\pm\infty$. (b) The nonadiabatic energy levels $E_{\pm}(t)$ of the model which
exhibit the real NLC at $t_c=0$.}
\label{figfield}
\end{figure}

To obtain the third protocol, we set $\theta (t)$ and $\varphi (t)$ as
\begin{equation}
\theta (t) =\arccos [\tanh (2\epsilon t)], ~~
\varphi (t) =\frac \pi 2-\arctan [\tanh (\epsilon t)],  \label{examp3}
\end{equation}
via which the driving field components are achieved as
\begin{equation}
\Omega _x(t) =2\epsilon \mbox{sech}(\epsilon t)\mbox{sech}^{\frac
12}(2\epsilon t), ~~
\Omega _z(t) =\epsilon [3\mbox{sech}(2\epsilon t)-2].  \label{hamil3z}
\end{equation}
The curves of the field pulses $\Omega _x(t)$ and $\Omega _z(t)$ are plotted
in Fig. \ref{figfield}(a). According to Eq. (\ref{examp3}), $\cos \varphi $ will evolve
from $-\frac{\sqrt{2}}2$ to $\frac{\sqrt{2}}2$ as the time goes from
$-\infty $ to $+\infty $. So the ratio $\cot \theta _h/\cot \theta $ should
alter its sign at $t\rightarrow \pm \infty $ according to Eq. (\ref{match}).
Specifically, the orientation of $\vec{\alpha}(t)$ will go from $-z$ to
$+z $, so the complete population transfer $|m\rangle\leftrightarrow |-m\rangle$
 can be realized through the nonadiabtic evolution. On the other hand,
as $\vec{\alpha}_h(t)$ tends to the $-z$ axis as $t\rightarrow
\pm \infty $, the adiabatic state will return to its initial state.
At $t=0$, there are $\theta (0)=\varphi (0)=\frac \pi 2$,
which indicates that $\vec{\alpha}(0)$ is along the $y$ axis, hence is
vertical with $\vec{\alpha}_h(t)$ at that time instant. In Fig. \ref{figfield}(b) we depict
the corresponding NLC for the model with $j=\frac 12$.

\section{Discussion and conclusion}

The anomalous dynamical evolution described above represents the
extreme behavior associated with the nonadiabatic driving.
It is worthy to mention that
such an extreme manifestation of the nonadiabtic effect could also be found in
the standard LZ model. According to the LZ formula, the nonadiabatic transition
between the two adiabatic states over the whole evolution
is specified by $P=e^{-\frac {\pi\Delta^2}{2\nu}}$ in which $\Delta$ describes
the constant coupling between the two bare states and $\nu$ stands for
the scanning rate of the linearly driving field.
If the scanning rate satisfies $\nu/\Delta^2\rightarrow +\infty$,
the nonadiabtic evolution will lead to completely opposite behavior with
that of the instantaneous adiabtic state. In comparison,
the driving fields in all our proposed protocols have finite scanning rate
hence are more realistic physically.

The driven model with the anomalous dynamical evolution offers an alternative
scenario for the state transfer that is distinctly different from the existing schemes,
e.g., the transitionless quantum driving \cite{berry}, the so-called shortcut-to-adiabaticity
protocol \cite{counter,short} and their recent development \cite{short2}.
Given a general time-dependent Hamiltonian $H(t)$,
the transitionless algorithm
cancels out the nonadiabatic effect by introducing an auxiliary counter-diabatic field
and ensures that the dynamical evolution of the system follows a normal trajectory,
i.e., remains in the instantaneous eigenstate of $H(t)$. In this sense, the strategies
of the anomalous dynamical evolution and of those existing protocols are complementary
and may have different applications to the design of quantum state control in experimental systems.

To summarize, we have investigated the anomalous dynamical evolution for the
time-dependent quantum systems by virtue of a reverse-engineering method.
Our study demonstrates that the occurring of the anomaly, i.e., the evolution
of the nonadiabatic states exhibits opposite behavior with that of the adiabatic ones,
is always concomitant with the NLC in these nonadiabtically driven systems.
In particular, we have proposed three cases of
such kind of driving protocols. In all these protocols the relative
orientation of the representative vectors of the adiabatic and nonadiabatic
bases undergoes a change from the
parallel state to the antiparallel state, and the nonadiabatic levels
of each system exhibit the anticipated crossing phenomenon.

\end{document}